\documentclass{article}
\usepackage{spconf, amsmath, graphicx, xcolor}
\usepackage{svg, adjustbox, subcaption, booktabs, amssymb, hyperref, fancyhdr}

\fancypagestyle{firstpage}
{ \fancyhead[L]{Submitted to IEEE ICIP 2024} 
 \fancyhead[R]{}} 

\newcommand{\reg}{\color{violet}}
\newcommand{\gan}{\color{blue}}
\newcommand{\flo}{\color{red}}

\title{Trustworthy SR: Resolving Ambiguity in Image Super-resolution \\ via Diffusion Models and Human Feedback}
\name{Cansu Korkmaz$^1$, Ege Cirakman$^2$, A.Murat Tekalp$^1$, Zafer Doğan$^1$ \thanks{This work was supported in part by an AI Fellowship to C. Korkmaz provided by the KUIS AI Center. This work was supported in part by TUBITAK 2247-A Award No.~120C156, TUBITAK 2232 Award No.~118C337, and KUIS AI Center funded by Turkish Is Bank. AMT acknowledges support from Turkish Academy of Sciences (TUBA).}}
\address{$^1$ College of Engineering and KUIS AI Center, 
Koç University, Istanbul, Turkey\\$^2$ Department of Electrical and Electronics Engineering, Istanbul Technical University, Istanbul, Turkey}

\begin{document}
%
\maketitle
\begin{abstract}
Super-resolution (SR) is an ill-posed inverse problem with a large set of feasible solutions that are consistent with a given low-resolution image. Various deterministic algorithms aim to find a single solution that balances fidelity and perceptual quality; however, this trade-off often causes visual artifacts that bring ambiguity in information-centric applications. On~the other hand, diffusion models (DMs) excel in generating a diverse set of feasible SR images that span the~solution space. The challenge is then how to determine the most likely solution among this set in a trustworthy manner. We  observe that quantitative measures, such as PSNR, LPIPS, DISTS, are not reliable indicators to resolve ambiguous cases. To this effect, we propose employing human feedback, where we ask  human subjects to select a small number of likely samples and we ensemble the averages of selected samples. This strategy leverages the high-quality image generation capabilities of DMs, while recognizing the importance of obtaining a single trustworthy solution, especially in use cases, such as identification of specific digits or letters, where generating multiple feasible solutions may not lead to a reliable outcome. Experimental results demonstrate that our proposed strategy provides more trustworthy solutions when compared to state-of-the art SR methods.
\end{abstract}

\begin{figure}[!t]
\centering
\begin{subfigure}{0.155\textwidth}
\centering
\captionsetup{justification=centering}
    \begin{subfigure}{\textwidth}
        \includegraphics[width=\textwidth]{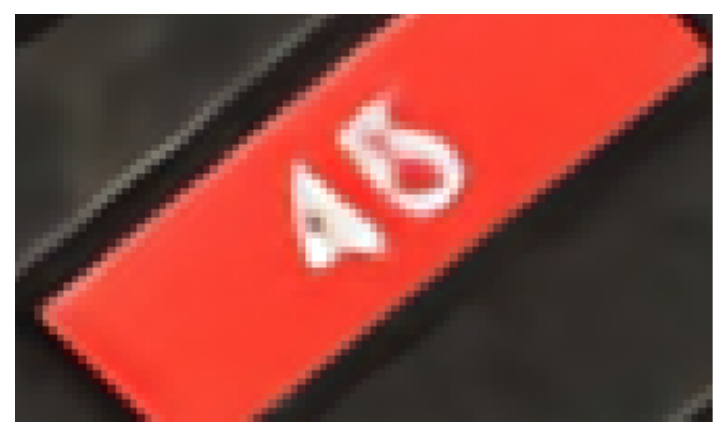}
    \end{subfigure} \vspace{-16pt}
    \caption*{\reg EDSR  \cite{EDSR2017}\\ (27.17 / 0.213)}
    \begin{subfigure}{\textwidth}
        \includegraphics[width=\textwidth]{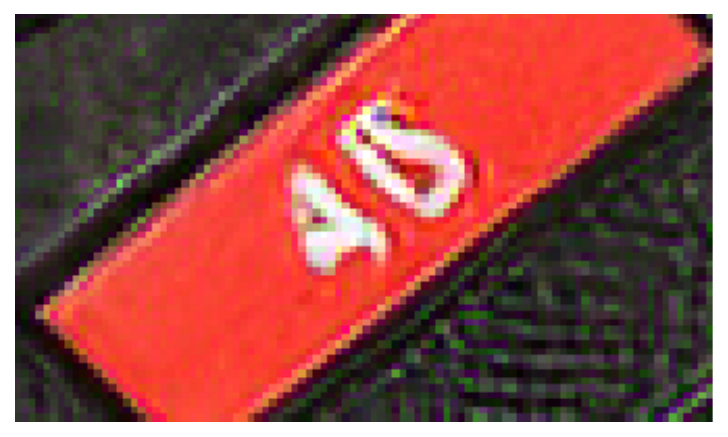}
    \end{subfigure} \vspace{-16pt}
    \caption*{\gan ESRGAN+ \cite{esrganplus} \\ (22.78 / 0.197)}
    \begin{subfigure}{\textwidth}
        \includegraphics[width=\textwidth]{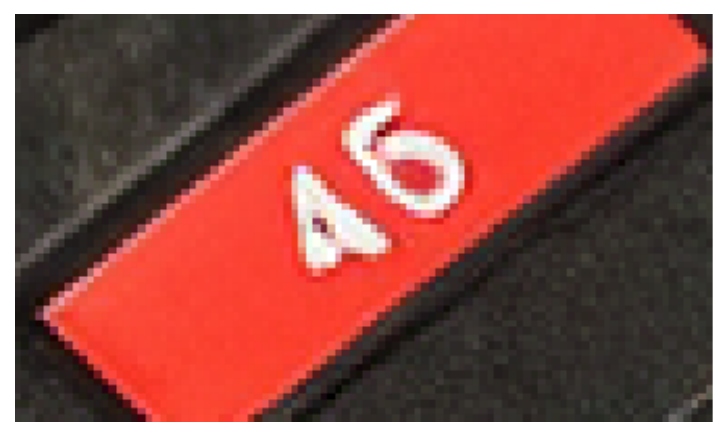}
    \end{subfigure} \vspace{-16pt}
    \caption*{\gan SROOE \cite{srooe_Park_2023_CVPR} \\ (26.67 / 0.184)}
\end{subfigure}
\begin{subfigure}{0.155\textwidth}
\centering
\captionsetup{justification=centering}
    \begin{subfigure}{\textwidth}
        \includegraphics[width=\textwidth]{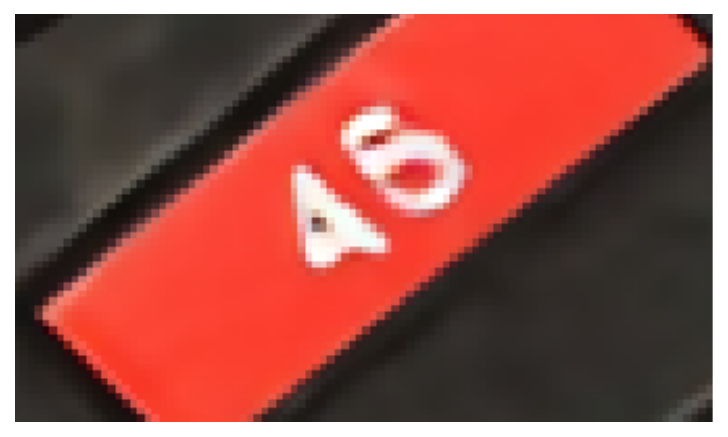}
    \end{subfigure} \vspace{-16pt}
    \caption*{\reg RRDB \cite{wang2018esrgan} \\ (26.46 / 0.223)}
    \begin{subfigure}{\textwidth}
        \includegraphics[width=\textwidth]{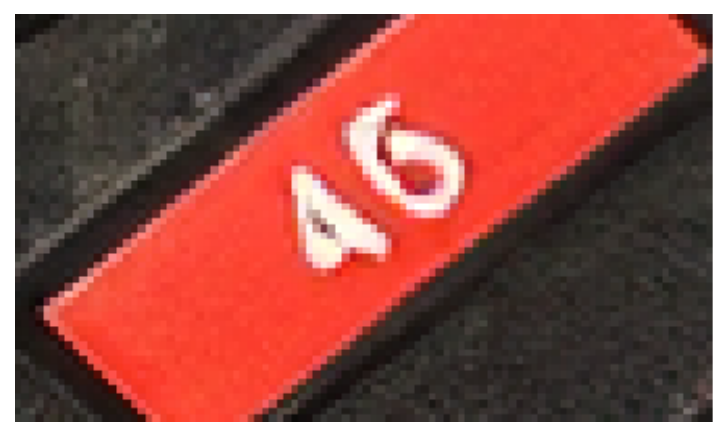}
    \end{subfigure} \vspace{-16pt}
    \caption*{\gan SPSR \cite{ma_SPSR} \\ (25.82 / 0.184)}
    \begin{subfigure}{\textwidth}
        \includegraphics[width=\textwidth]{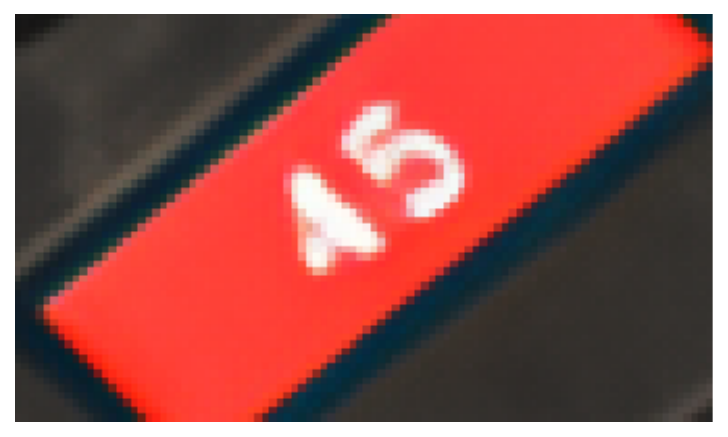}
    \end{subfigure} \vspace{-16pt}
    \caption*{LDM-SS (Ours) \\ (26.69 / 0.193)}
\end{subfigure}
\begin{subfigure}{0.155\textwidth}
\centering
\captionsetup{justification=centering}
     \begin{subfigure}{\textwidth}
        \includegraphics[width=\textwidth]{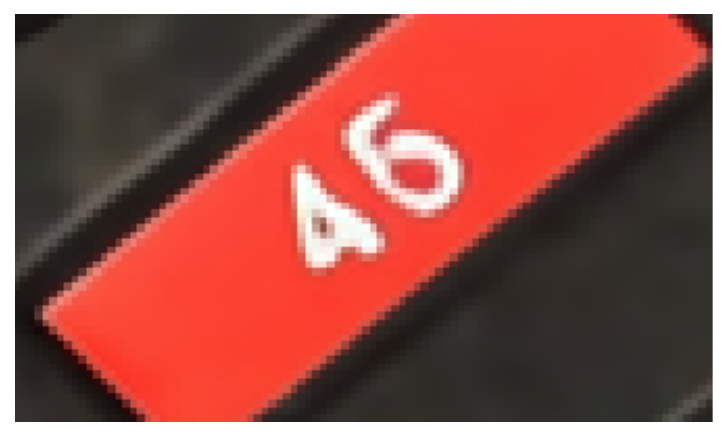}
    \end{subfigure} \vspace{-16pt}
    \caption*{\reg HAT \cite{hat_chen2023activating} \\ (28.03 / 0.217)}
    \begin{subfigure}{\textwidth}
        \includegraphics[width=\textwidth]{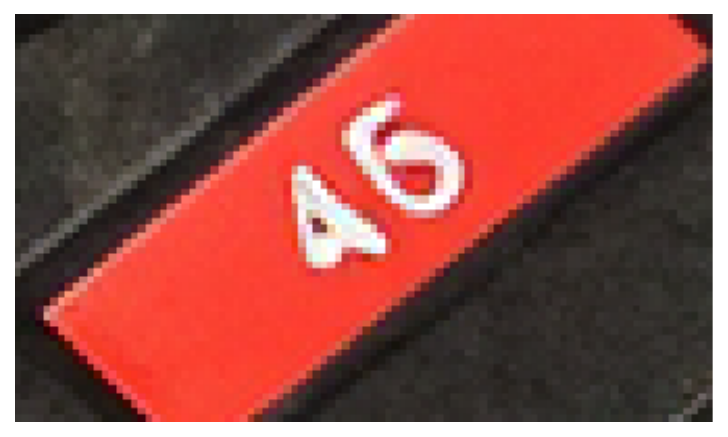}
    \end{subfigure} \vspace{-16pt}
    \caption*{\gan LDL \cite{details_or_artifacts} \\ (26.67 / 0.194)}
    \begin{subfigure}{\textwidth}
        \includegraphics[width=\textwidth]{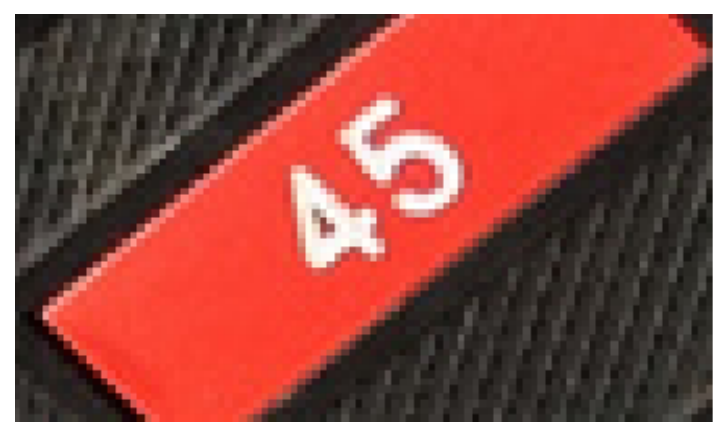}
    \end{subfigure} \vspace{-16pt}
    \caption*{HR \\ (PSNR$\uparrow$ / DISTS$\downarrow$)}
\end{subfigure} 
\caption{Visual performance of recent $\times$4 SR methods on a~crop from Urban100 dataset (img-6) \cite{urban100_cite}. SOTA methods reconstruct ``5" as ``6", whereas the opening in the lower part of ``5" is visible in our result confirming that the proposed strategy resolves the ambiguity and provide a thrustworthy solution. Note PSNR, DISTS and other quantitative scores are not reliable indicators to resove such ambiguity.}
\label{fig:first_img} 
\end{figure}

\section{Introduction}
\label{sec:intro}

Single-image super-resolution (SR) is an ill-posed inverse problem, where a single input image can correspond to multiple feasible output images, introducing ambiguity into the~SR reconstruction process. Early deep learning based SR methods~\cite{dong2015image, kim2016accurate, ledig2017photorealistic, EDSR2017, srcnn, RCAN2018, wang2018esrgan} treated SR as a deterministic regression problem and generated a single output image based on a set of LR-HR paired training data. 

Recognizing the inherently ill-posed nature of the SR problem, many recent methods~\cite{lugmayr2020srflow, jo2021srflowda, SR3_saharia2022image, LDM_rombach2022high, korkmaz2022perception, diff_mixture_of_experts_luo2023image} and challenges~\cite{learning_SR_space_2021, learning_SR_space_2022} propose learning a one-to-many mapping that is consistent with the conditional distribution of output images given the input, aiming to generate a diverse set of feasible SR images that span the SR space. These formulations prioritize photo-realism of solutions, consistency with the LR image, and diversity (coverage of the SR~space). However, such stochastic generative models introduce a new problem: how to find a trustworthy solution when extracting critical information from many possibly conflicting SR images, e.g.,~whether a digit is 5 or 6. In such cases, there~is need for reliable methodology to determine whether it is possible to rule out certain solutions and pick a trustworthy outcome.

We observe that the traditional objective quality (fidelity) measures such as Peak Signal-to-Noise Ratio (PSNR) and SSIM, as well as popular perceptual quality measures such as LPIPS~\cite{lpips} and DISTS \cite{dists}, are inadequate to evaluate trustworthy SR solutions. Indeed there exists a divergence between human visual evaluations and known quantitative measures, particularly when evaluating generated high-frequency~(HF) details.  Therefore, one cannot rely solely on numerical measures to evaluate trustworthiness in SR.   

To this effect, we propose a human feedback-centric approach to assess trustworthiness of SR solutions. Human subjects are asked to select up to 5 most likely outputs from a set of feasible SR images. The images selected by each subject are then ensembled according to the requirements of the task. We introduce a straightforward yet highly effective image ensembling strategy in this study, enabling  diffusion models to leverage human feedback to resolve the ambiguity in generated samples. In essence, our goal is to address the challenge of finding a trustworthy solution with accurate details. Our experimental results demonstrate that a~pre-trained Latent Diffusion Model (LDM) with strategic sample selection guided by human feedback (LDM-SS) and image ensembling outperforms state-of-the-art SR methods considering both reliability and visual image quality. However, we observe that the success of LDM-SS in improving trustworthiness and subjective visual quality does not necessarily translate into improvements in traditional quantitative measures.

Our main contributions are summarized as follows:\\
1. We address the challenge of obtaining a single trustworthy solution in the SR space spanned by diffusion models when information extraction is critical. \\
2. We introduce a human feedback-centric approach for SR sample selection, along with an ensembling strategy, demonstrating the superiority of the diffusion model in achieving accurate and reliable results. \\
3. Our suggested approach is versatile, as the combination of ensembling strategy and the integration of human feedback can be applied to any stochastic generative model, guiding it to produce trustworthy and consistent SR images. 

\section{Related Works}
\label{sec:related_works}

\noindent \textbf{One-to-One SR Inference.} 
Many popular SR models, including 
EDSR \cite{EDSR2017}, RRDB~\cite{wang2018esrgan} and PDASR \cite{PDASR}, are deterministic regressive mappings from LR to HR images trained by $l_1$ or $l_2$ reconstruction losses. Even though these models achieve high fidelity, measured by PSNR, they still produce serious artifacts that contribute to the ambiguity problem. 

Generative adversarial networks (GAN) have been proposed to generate photo-realistic images~\cite{ian_gan}. Many SR models based on the principles of GAN have been proposed over the years~\cite{ledig2017photorealistic}, \cite{wang2018esrgan}, \cite{esrganplus}, \cite{ma_SPSR}, \cite{details_or_artifacts},~\cite{srooe_Park_2023_CVPR} to generate a single SR image (per~$\lambda$).
It is well known that GANs hallucinate HF details. It is obvious to humans that some of~these hallucinations are artifacts, while some others may look~like real although they are fake. Hence, GAN-based SR~models cannot offer a~trustworthy solution to resolve the ambiguity~problem.
\vspace{-8pt}

\noindent \textbf{One-to-Many SR Inference.}
To generate rich diversity, likelihood-based model training prioritizing accurate density estimation such as variational autoencoders (VAE)~\cite{vae_liu2021variational, vae_zhou2021vspsr} and normalizing flow based SR methods \cite{lugmayr2020srflow, jo2021srflowda, fsncsr_song2022fs} have been introduced. They offer notable benefits compared to GAN-based methods including monotonic convergence, stable training and efficiency while generating multiple SR images, however, they exhibit low fidelity in terms of image quality. Similarly, autoregressive models (ARM) \cite{arm_van2016conditional, arm_pmlr-v119-chen20s} excel in density estimation but face high computational complexity for inference due to their sequential sampling process. In addition, pixel-based image representations require prolonged training times to learn subtle HF details~\cite{arm_salimans2017pixelcnn++}. 

Recently, significant progress has been made in one-to-many SR image generation with the advent of diffusion models \cite{diff_NEURIPS2020_4c5bcfec, diff_implicit_chen2021learning, diff_avrahami2022blended, diff_gu2022vector, diffretrival_NEURIPS2022_62868cc2, SR3_saharia2022image, LDM_rombach2022high, diff_mixture_of_experts_luo2023image}. 
However, current diffusion models still face some challenges, including but not limited to complex two-stage pipelines, high computational requirements for training, and presence of unnatural artifacts resulting in unreliable ambiguous SR outputs. In this work, we employ a pre-trained LDM for $\times$4 SR to avoid lengthy training and propose strategic sample selection via human feedback to address the challenge of trustworty SR image reconstruction by integrating the strengths of diffusion models, human feedback, and image ensembling within a practical framework.


\begin{figure}[b!]
\centering
\includegraphics[width=\linewidth]{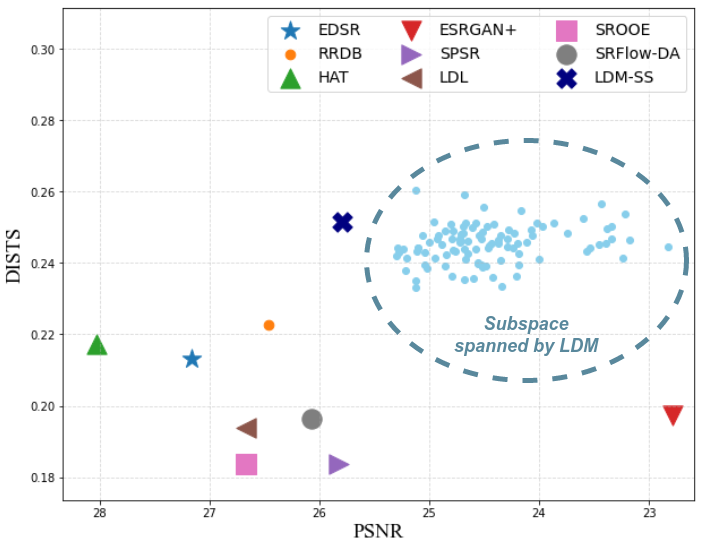}
\caption{Demonstration of the SR space spanned by LDM \cite{LDM_rombach2022high} samples, proposed LDM-SS and other state-of-the-art methods on the PSNR-DISTS plane. We note that perception-distortion tradeoff with respect to known metrics does not correlate well with visual quality and trustworthiness.}
\label{fig:pd_tradeoff}
\end{figure}

\begin{figure*}[t!]
\centering
\includegraphics[width=\linewidth]{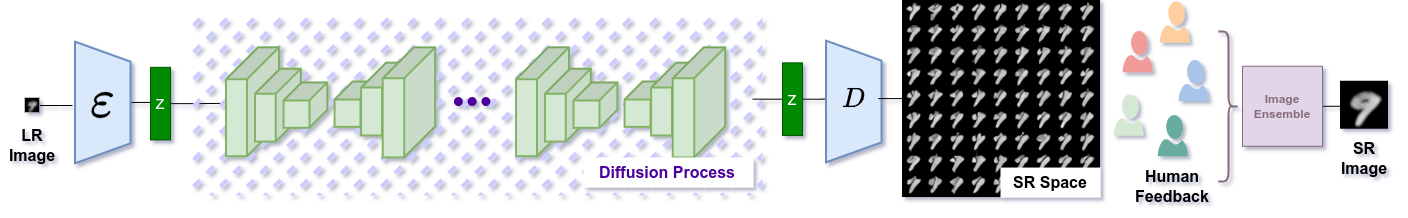} \vspace{-20pt}
\caption{Block diagram of our approach depicting 
sample selection from the diffusion model SR space by human~feedback. 
}
\label{fig:arch}
\end{figure*}

\section{LDM-SS: Resolving Ambiguity through Sample Selection in Diffusion SR Space}
\label{sec:method}

\subsection{SR Space Spanned by LDM}
Diffusion models \cite{diff_NEURIPS2020_4c5bcfec} are statistical models designed to learn a data distribution, $p(x)$, through progressively denoising a normal distributed variable. This process involves learning the inverse operation of a Markov Chain with a fixed length~$T$. Recently proposed LDM \cite{LDM_rombach2022high} method performs diffusion process in a low dimensional latent space and provides computationally tractable and flexible SR images what we refer as subspace spanned by LDM as demonstrated in Fig. \ref{fig:pd_tradeoff}. On~one hand all diffusion-based methods including LDM have a common problem: generation of SR image samples exhibiting rich textures but lack fidelity. On the other hand, in order to compare the performance of DMs and one-to-one SR approaches, we need to map the set of outputs to a single SR image. One can simply select a certain realization from the set. However, since there is a significant variety among all possible SR realizations produced by DMs, it is hard to obtain the well-suited result in the first realization. Therefore, generating numerous possible solutions in such information-centric applications may not result in a conclusive decision.  

\subsection{Resolving Ambiguity in the Diffusion SR Space}
Our proposed method involves combining various diffusion samples into a unified, trustworthy image by ensembling them through human feedback. The objective is to strike a balance between high fidelity and perceptual distortion, particularly for information-centric applications.
The block diagram illustrating our proposed approach is depicted in Fig. \ref{fig:arch}. Specifically, while developing the concept of strategic sample selection for trustworthy SR images, we employ LDM method to construct SR space by sampling from the learned distribution at inference time. For each LR image, we generate up to 100 images for human selection, then top 5 selected images by majority voting are ensembled by pixel-wise averaging to construct the SR image. 

\begin{table}[t]
\caption{Thirty participants were asked to select two samples out of 324 generated samples that are most helpful to identify the specific digit.}   \vspace{-18pt}
\begin{center}
\begin{adjustbox}{width=0.4\textwidth}
\begin{tabular}{lll}
\toprule
 & Perceived as ``5" & Perceived as ``6" \\
\midrule
$\#$ of People & 22 (73.3\%) & 8 (26.7\%) \\
\bottomrule
\end{tabular}
\end{adjustbox}
\label{table:5or6_results}
\end{center}  
\end{table}

\begin{figure}[h]
\centering
\begin{subfigure}{0.03\textwidth}
     \begin{subfigure}{\textwidth}
        \includegraphics[width=\textwidth]{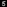} 
    \end{subfigure}
    \caption*{LR}
\end{subfigure}
\begin{subfigure}{0.1\textwidth}
     \begin{subfigure}{\textwidth}
        \includegraphics[width=\textwidth]{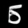} 
    \end{subfigure}
    \caption*{\reg HAT\cite{hat_chen2023activating}}
\end{subfigure}
\begin{subfigure}{0.1\textwidth}
    \begin{subfigure}{\textwidth}
        \includegraphics[width=\textwidth]{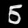} 
    \end{subfigure}
\caption*{\gan SROOE\cite{srooe_Park_2023_CVPR}}
\end{subfigure}
\begin{subfigure}{0.1\textwidth}
     \begin{subfigure}{\textwidth}
        \includegraphics[width=\textwidth]{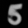} 
    \end{subfigure}
    \caption*{Avg. ``5"}
\end{subfigure}
\begin{subfigure}{0.1\textwidth}
    \begin{subfigure}{\textwidth}
        \includegraphics[width=\textwidth]{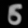} 
    \end{subfigure}
\caption*{Avg. ``6"}
\end{subfigure}
\caption{Identification of the digit from LR image is impossible and results of state-of-the-art methods HAT \cite{hat_chen2023activating} (Regressive) and SROOE \cite{srooe_Park_2023_CVPR} (GAN-SR) are ambiguous. However, the five most selected figures were combined through pixel-wise averaging, yielding single, informative SR image. The prevalence of the perception of ``5" enables mitigating ambiguity.}
\label{fig:5or6} 
\end{figure}

\begin{figure*}[ht]
\centering
\begin{subfigure}{0.075\textwidth}
     \begin{subfigure}{\textwidth}
        \includegraphics[width=\textwidth]{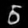} \\ \footnotesize \reg EDSR \\ \cite{EDSR2017} \\ 20.28 \\ 0.146
    \end{subfigure}
     \begin{subfigure}{\textwidth}
        \includegraphics[width=\textwidth]{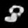} \\ \footnotesize \reg EDSR \\ \cite{EDSR2017} \\ 18.74 \\ 0.136
    \end{subfigure}
\end{subfigure}
\begin{subfigure}{0.075\textwidth}
    \begin{subfigure}{\textwidth}
        \includegraphics[width=\textwidth]{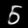} \\ \footnotesize \reg RRDB \\ \cite{wang2018esrgan} \\ 20.99 \\ 0.102
    \end{subfigure}
    \begin{subfigure}{\textwidth}
        \includegraphics[width=\textwidth]{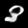} \\ \footnotesize \reg RRDB \\ \cite{wang2018esrgan} \\ 18.43 \\ 0.146
    \end{subfigure}
\end{subfigure}
\begin{subfigure}{0.075\textwidth}
    \begin{subfigure}{\textwidth}
        \includegraphics[width=\textwidth]{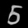} \\ \footnotesize \reg HAT \\ \cite{hat_chen2023activating} \\ 20.46 \\ 0.125
    \end{subfigure}
    \begin{subfigure}{\textwidth}
        \includegraphics[width=\textwidth]{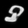} \\\footnotesize \reg HAT \\ \cite{hat_chen2023activating} \\ 17.96 \\ 0.136
    \end{subfigure}
\end{subfigure}
\begin{subfigure}{0.075\textwidth}
    \begin{subfigure}{\textwidth}
        \includegraphics[width=\textwidth]{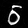} \\ \footnotesize \gan ESRGAN+ \cite{esrganplus} \\ 19.00 \\ 0.108
    \end{subfigure}
    \begin{subfigure}{\textwidth}
        \includegraphics[width=\textwidth]{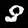} \\ \footnotesize \gan ESRGAN+ \cite{esrganplus} \\ 16.61 \\ 0.122
    \end{subfigure}
\end{subfigure}
\begin{subfigure}{0.075\textwidth}
    \begin{subfigure}{\textwidth}
        \includegraphics[width=\textwidth]{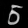} \\ \footnotesize \gan SPSR \\ \cite{ma_SPSR} \\ 19.87 \\ 0.125
    \end{subfigure}
    \begin{subfigure}{\textwidth}
        \includegraphics[width=\textwidth]{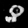} \\ \footnotesize \gan SPSR \\ \cite{ma_SPSR} \\ 16.90 \\ 0.145
    \end{subfigure}
\end{subfigure}
\begin{subfigure}{0.075\textwidth}
    \begin{subfigure}{\textwidth}
        \includegraphics[width=\textwidth]{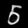} \\ \footnotesize \gan LDL \\ \cite{details_or_artifacts} \\21.29 \\ 0.095
    \end{subfigure}
    \begin{subfigure}{\textwidth}
        \includegraphics[width=\textwidth]{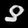} \\ \footnotesize \gan LDL \\ \cite{details_or_artifacts} \\17.84 \\ 0.118
    \end{subfigure}
\end{subfigure} 
\begin{subfigure}{0.075\textwidth}
    \begin{subfigure}{\textwidth}
        \includegraphics[width=\textwidth]{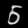} \\ \footnotesize \gan SROOE \\ \cite{srooe_Park_2023_CVPR} \\ 20.93 \\ 0.106
    \end{subfigure}
    \begin{subfigure}{\textwidth}
        \includegraphics[width=\textwidth]{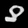} \\ \footnotesize \gan SROOE \\ \cite{srooe_Park_2023_CVPR} \\ 18.37 \\ 0.125
    \end{subfigure}
\end{subfigure}
\begin{subfigure}{0.075\textwidth}
    \begin{subfigure}{\textwidth}
        \includegraphics[width=\textwidth]{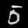} \\ \footnotesize \flo HCFlow++ \cite{hcflow_liang21hierarchical} \\ 19.45 \\ 0.121
    \end{subfigure}
    \begin{subfigure}{\textwidth}
        \includegraphics[width=\textwidth]{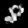} \\ \footnotesize \flo HCFlow++ \cite{hcflow_liang21hierarchical} \\ 17.85 \\ 0.204
    \end{subfigure}
\end{subfigure}
\begin{subfigure}{0.075\textwidth}
    \begin{subfigure}{\textwidth}
        \includegraphics[width=\textwidth]{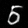} \\ \footnotesize \flo SRFlowDA \cite{jo2021srflowda} \\ 21.33 \\ 0.099
    \end{subfigure}
    \begin{subfigure}{\textwidth}
        \includegraphics[width=\textwidth]{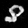} \\ \footnotesize \flo SRFlowDA \cite{jo2021srflowda} \\ 17.68 \\ 0.151
    \end{subfigure}
\end{subfigure}
\begin{subfigure}{0.075\textwidth}
    \begin{subfigure}{\textwidth}
        \includegraphics[width=\textwidth]{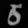} \\ \footnotesize LDM \\ \cite{LDM_rombach2022high} \\ 16.89 \\ 0.212
    \end{subfigure}
    \begin{subfigure}{\textwidth}
        \includegraphics[width=\textwidth]{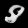} \\ \footnotesize LDM \\ \cite{LDM_rombach2022high} \\ 16.73 \\ 0.190
    \end{subfigure}
\end{subfigure}
\begin{subfigure}{0.075\textwidth}
    \begin{subfigure}{\textwidth}
        \includegraphics[width=\textwidth]{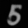} \\ \footnotesize LDM-SS \\ (Ours) \\ 17.62 \\ 0.215
    \end{subfigure}
    \begin{subfigure}{\textwidth}
        \includegraphics[width=\textwidth]{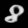} \\ \footnotesize LDM-SS \\ (Ours) \\ 17.07 \\ 0.151
    \end{subfigure}
\end{subfigure}
\begin{subfigure}{0.075\textwidth}
    \begin{subfigure}{\textwidth}
        \includegraphics[width=\textwidth]{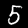} \\ \footnotesize HR \\ \cite{deng2012mnist}\\ PSNR$\uparrow$ \\ DISTS$\downarrow$\cite{dists}
     \end{subfigure}
    \begin{subfigure}{\textwidth}
        \includegraphics[width=\textwidth]{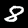} \\ \footnotesize HR \\ \cite{deng2012mnist}\\ PSNR$\uparrow$ \\ DISTS$\downarrow$\cite{dists}
    \end{subfigure}
\end{subfigure} 
\caption{Visual comparison of proposed method and state-of-the art regressive (purple), GAN-based (blue), flow-based (red) and LDM SR methods on Mnist dataset \cite{deng2012mnist}. It can be seen that the proposed LDM-SS method provides more reliable SR images for information retrieval, but quantitative metrics are insufficient to capture the nuances of visual artifacts or trustworthiness.}
\label{fig:mnist} 
\end{figure*}

A hypothetical use case is demonstrated in the following example: Suppose an evidence, shown in Fig. \ref{fig:5or6}-(a), is presented to a~court of law. The prosecution confidently asserted that the~digit in the image is unequivocally ``5." On the opposing side, the defense argued persuasively for an undeniable ``6." As experts in image processing were summoned to the witness stand, they face a tough challenge. They have employed state-of-the-art regressive and generative SR algorithms, including HAT \cite{hat_chen2023activating} and SROOE \cite{srooe_Park_2023_CVPR}. Yet, they have not reached a consensus. This ambiguity in SR problems  exemplifies the profound complexities awaiting a solution. Since~the correct answer, the presented digit being 5 or 6 can be deduced from the plausible SR space spanned by diffusion samples, then direct human evaluation can be employed. Specifically, we use mean opinion score (MOS) and total of 30 participants are tasked with identifying the specific number depicted. Subsequently, they are asked to select the 2 samples among 324 generated ones that are most helpful to identify the digit. As presented in Table \ref{table:5or6_results}, 22 among 30 participants perceived generated samples as ``5", whereas 8 of them predicted the number as ``6". Then, 5 most selected figures for the digit is ensembled by pixel-wise averaging resulting in a single, informative SR image demonstrated in Fig.~\ref{fig:5or6}-(d) and (e). It is important to note that participants are not provided with information regarding the actual, ground-truth digit, relying solely on their visual preferences. Since, most people perceived the number as ``5", the ambiguity is resolved to sum extend.

\section{Experimental Results}
\label{sec:experimental_results}
\textbf{Implementation Details and Benchmarks.}
We selected two popular datasets as benchmarks: Mnist dataset \cite{deng2012mnist} and DIV2K \cite{Agustsson_2017_CVPR_Workshops}. Since, the size of Mnist images is 28x28, we obtain 7x7 LR images after downsampling by 4 in each dimension using Matlab's bicubic kernel. Then, each LR image is upsampled by the latent diffusion model (LDM) \cite{LDM_rombach2022high}. Since LDM is designed to upsample images from 64x64 to 256x256, 7x7 LR sample is repeated 9 times both vertically and horizontally before applying the diffusion process. Since~it~is a stochastic process each SR image contains variety of upsampled numbers. Similarly, cropped 32x32 pixels of RGB images from DIV2K \cite{Agustsson_2017_CVPR_Workshops} are fed into the pretrained LDM model with a scaling factor of 4$\times$. 

\begin{figure*}
\centering
\begin{subfigure}{0.16\textwidth}
     \begin{subfigure}{\textwidth}
        \includegraphics[width=\textwidth]{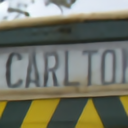} \\ \small \reg EDSR \cite{EDSR2017} \\ (31.88 / 0.137)
    \end{subfigure}
    \begin{subfigure}{\textwidth}
        \includegraphics[width=\textwidth]{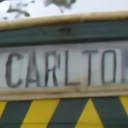} \\ \small \reg RRDB \cite{wang2018esrgan} \\ (30.51 / 0.121)
    \end{subfigure}
     \begin{subfigure}{\textwidth}
        \includegraphics[width=\textwidth]{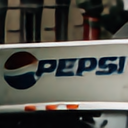} \\ \small \reg EDSR \cite{EDSR2017}\\ (23.88 / 0.213)
    \end{subfigure}
    \begin{subfigure}{\textwidth}
        \includegraphics[width=\textwidth]{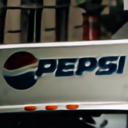} \\ \small \reg RRDB \cite{wang2018esrgan} \\ (23.81 / 0.214)
    \end{subfigure}
     \begin{subfigure}{\textwidth}
        \includegraphics[width=\textwidth]{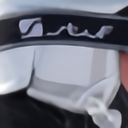} \\ \small \reg EDSR \cite{EDSR2017}\\ (27.54 / 0.211)
    \end{subfigure}
    \begin{subfigure}{\textwidth}
        \includegraphics[width=\textwidth]{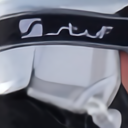} \\ \small \reg RRDB \cite{wang2018esrgan} \\ (27.07 / 0.196)
    \end{subfigure}
\end{subfigure}
\begin{subfigure}{0.16\textwidth}
    \begin{subfigure}{\textwidth}
        \includegraphics[width=\textwidth]{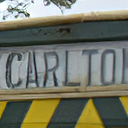} \\ \small \gan ESRGAN+ \cite{esrganplus} \\ (27.09 / 0.140)
    \end{subfigure}
    \begin{subfigure}{\textwidth}
        \includegraphics[width=\textwidth]{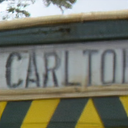} \\ \small \gan SPSR \cite{ma_SPSR} \\ (31.61 / 0.100)
    \end{subfigure}
    \begin{subfigure}{\textwidth}
        \includegraphics[width=\textwidth]{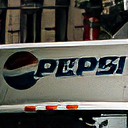} \\ \small \gan ESRGAN+ \cite{esrganplus} \\ (21.35 / 0.169)
    \end{subfigure}
    \begin{subfigure}{\textwidth}
        \includegraphics[width=\textwidth]{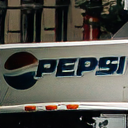} \\ \small \gan SPSR \cite{ma_SPSR} \\ (22.98 / 0.127)
    \end{subfigure}
    \begin{subfigure}{\textwidth}
        \includegraphics[width=\textwidth]{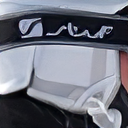} \\ \small \gan ESRGAN+ \cite{esrganplus} \\ (24.59 / 0.188)
    \end{subfigure}
    \begin{subfigure}{\textwidth}
        \includegraphics[width=\textwidth]{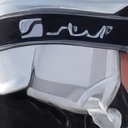} \\ \small \gan SPSR \cite{ma_SPSR} \\ (25.96 / 0.174)
    \end{subfigure}
\end{subfigure}
\begin{subfigure}{0.16\textwidth}
    \begin{subfigure}{\textwidth}
        \includegraphics[width=\textwidth]{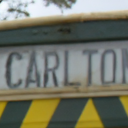} \\ \small \gan LDL \cite{details_or_artifacts} \\ (32.39 / 0.100)
    \end{subfigure}
    \begin{subfigure}{\textwidth}
        \includegraphics[width=\textwidth]{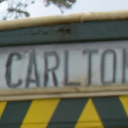} \\ \small \gan SROOE \cite{srooe_Park_2023_CVPR} \\ (33.06 / 0.085)
    \end{subfigure}
    \begin{subfigure}{\textwidth}
        \includegraphics[width=\textwidth]{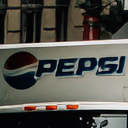} \\ \small \gan LDL \cite{details_or_artifacts} \\ (23.84 / 0.134)
    \end{subfigure}
    \begin{subfigure}{\textwidth}
        \includegraphics[width=\textwidth]{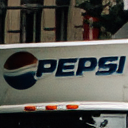} \\ \small \gan SROOE \cite{srooe_Park_2023_CVPR} \\ (24.60 / 0.122)
    \end{subfigure}
    \begin{subfigure}{\textwidth}
        \includegraphics[width=\textwidth]{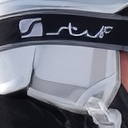} \\ \small \gan LDL \cite{details_or_artifacts} \\ (27.09 / 0.165)
    \end{subfigure}
    \begin{subfigure}{\textwidth}
        \includegraphics[width=\textwidth]{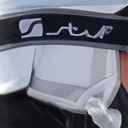} \\ \small \gan SROOE \cite{srooe_Park_2023_CVPR} \\ (27.76 / 0.158)
    \end{subfigure}
\end{subfigure}
\begin{subfigure}{0.16\textwidth}
    \begin{subfigure}{\textwidth}
        \includegraphics[width=\textwidth]{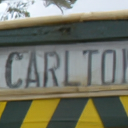} \\\small \flo HCFlow++ \cite{hcflow_liang21hierarchical} \\ (30.65 / 0.116)
    \end{subfigure}
    \begin{subfigure}{\textwidth}
        \includegraphics[width=\textwidth]{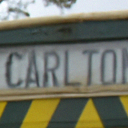} \\\small \flo SRFlow-DA \cite{jo2021srflowda} \\ (32.14 / 0.116)
    \end{subfigure}
    \begin{subfigure}{\textwidth}
        \includegraphics[width=\textwidth]{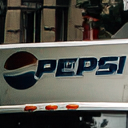} \\\small  \flo HCFlow++ \cite{hcflow_liang21hierarchical} \\ (23.40 / 0.141)
    \end{subfigure}
    \begin{subfigure}{\textwidth}
        \includegraphics[width=\textwidth]{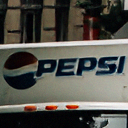} \\\small \flo SRFlow-DA \cite{jo2021srflowda} \\ (24.81 / 0.143)
    \end{subfigure}
    \begin{subfigure}{\textwidth}
        \includegraphics[width=\textwidth]{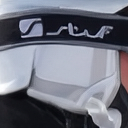} \\\small  \flo HCFlow++ \cite{hcflow_liang21hierarchical} \\ (27.24 / 0.161)
    \end{subfigure}
    \begin{subfigure}{\textwidth}
        \includegraphics[width=\textwidth]{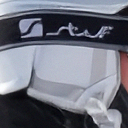} \\\small \flo SRFlow-DA \cite{jo2021srflowda} \\ (27.15 / 0.154)
    \end{subfigure}
\end{subfigure}
\begin{subfigure}{0.16\textwidth}
    \begin{subfigure}{\textwidth}
        \includegraphics[width=\textwidth]{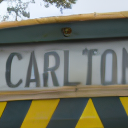} \\ \small LDM \cite{LDM_rombach2022high} \\ (28.49 / 0.169)
    \end{subfigure}
    \begin{subfigure}{\textwidth}
        \includegraphics[width=\textwidth]{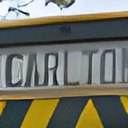} \\ \small IDM \cite{implicit_gao2023implicit} \\ (24.04 / 0.205)
    \end{subfigure}
    \begin{subfigure}{\textwidth}
        \includegraphics[width=\textwidth]{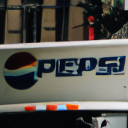} \\ \small  LDM \cite{LDM_rombach2022high}  \\ (22.11 / 0.179)
    \end{subfigure}
    \begin{subfigure}{\textwidth}
        \includegraphics[width=\textwidth]{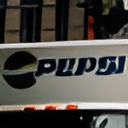} \\ \small  IDM \cite{implicit_gao2023implicit} \\ (21.13 / 0.212)
    \end{subfigure}
    \begin{subfigure}{\textwidth}
        \includegraphics[width=\textwidth]{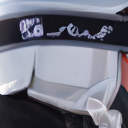} \\ \small  LDM \cite{LDM_rombach2022high} \\ (23.99 / 0.195)
    \end{subfigure}
    \begin{subfigure}{\textwidth}
        \includegraphics[width=\textwidth]{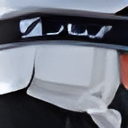} \\ \small IDM \cite{implicit_gao2023implicit} \\ (23.24 / 0.227)
    \end{subfigure}
\end{subfigure}
\begin{subfigure}{0.16\textwidth}
    \begin{subfigure}{\textwidth}
        \includegraphics[width=\textwidth]{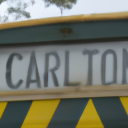} \\ \small LDM-SS (Ours) \\(26.71 / 0.209)
    \end{subfigure}
        \begin{subfigure}{\textwidth}
        \includegraphics[width=\textwidth]{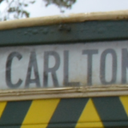} \\ \small HR (img-832)\\ (PSNR$\uparrow$/DISTS$\downarrow$\cite{dists})
    \end{subfigure}
    \begin{subfigure}{\textwidth}
        \includegraphics[width=\textwidth]{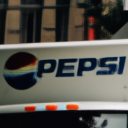} \\ \small LDM-SS (Ours) \\(23.98 / 0.242)
    \end{subfigure}
        \begin{subfigure}{\textwidth}
        \includegraphics[width=\textwidth]{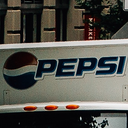} \\ \small HR (img-861)\\ (PSNR$\uparrow$/DISTS$\downarrow$\cite{dists})
    \end{subfigure}
    \begin{subfigure}{\textwidth}
        \includegraphics[width=\textwidth]{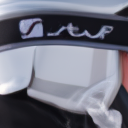} \\ \small LDM-SS (Ours) \\(25.92 / 0.227)
    \end{subfigure}
        \begin{subfigure}{\textwidth}
        \includegraphics[width=\textwidth]{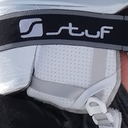} \\ \small HR (img-894)\\ (PSNR$\uparrow$/DISTS$\downarrow$\cite{dists})
    \end{subfigure}
\end{subfigure} 
\caption{Comparison of the proposed method with the state-of-the-art for $\times$4 SR on natural images from DIV2K validation set~\cite{Agustsson_2017_CVPR_Workshops}. Even though the proposed LDM-SS method with human feedback has clear advantages in reconstructing realistic high-frequency details while inhibiting artifacts, popular quantitative metrics are insufficient to reflect the visual improvements.}
\label{fig:visuals_div2k} 
\end{figure*}


\noindent
\textbf{Human Evaluation.}
Since none of the highly utilized evaluation metrices including PSNR, LR-Consistency \cite{learning_SR_space_2021}, SSIM, LPIPS \cite{lpips} and DISTS \cite{dists} do not correlate with the information that has been conveyed through the SR images. Accordingly, we employ direct human assessment for evaluation. While the mean opinion score (MOS) is a prevalent measure for assessing image quality has been found to be a more reliable method for such subjective quality assessments. In Task-1, the 30 participants are assigned the task of identifying a specific number from the Mnist \cite{deng2012mnist} dataset, followed by the requirement to select the two numbers deemed most "natural" from a pool of 324 generated samples. Task-2 is similar, subjects are presented 15 natural images from DIV2K \cite{Agustsson_2017_CVPR_Workshops}, focusing solely on selecting the more photo-realistic image. In both tasks, the ground-truth image is not provided, hence participants just rely entirely on their visual preferences.

\begin{table*}[t]
\begin{center}
\begin{adjustbox}{width=0.95\textwidth}
\begin{tabular}{lccccccccc}
\toprule
 & SR Model & PSNR$\uparrow$ & LR Consistency $\uparrow$ & SSIM$\uparrow$ & LPIPS & LPIPS$_{\text{VGG}}$$\downarrow$ & PieAPP$\downarrow$ & DISTS$\downarrow$ & NRQM$\uparrow$ \\
\midrule
& \reg EDSR &  25.962 & 43.047 & 0.803 &  0.115 &  0.231 &  0.901 &  0.194 &  5.142 \\
\reg Regressive & \reg RRDB &  25.316 & 39.508 & 0.788 &  0.103 &  0.225 &  0.799 &  0.187 &  5.850 \\
 & \reg HAT &  27.408 & 44.673 & 0.826 &  0.089 &  0.201 &  0.686 &  0.179 &  5.518 \\ \hline
& \gan ESRGAN+ &  22.666 &  31.718 & 0.716 &  0.083 &  0.224 &  0.292 &  0.168 &  7.757 \\
\gan GAN-based & \gan SPSR &  24.760 & 36.520 & 0.762 &  0.063 &  0.184 &  0.523 &  0.138 &  7.159 \\ 
& \gan LDL &  27.194 & 43.360 & 0.852 &  0.053 &  0.145 &  0.396 &  0.125 &  7.079 \\ 
& \gan SROOE &  25.894 & 41.040 & 0.790 &  0.061 &  0.166 &  0.562 &  0.132 &  6.741 \\ \hline
\flo Flow-based & \flo SRFlowDA &  27.510 & 46.929 & 0.852 &  0.062 &  0.172 &  0.686 &  0.145 &  6.699 \\ 
& \flo HCFlow &  25.062 & 43.302 & 0.777 &  0.067 &  0.183 &  0.641 &  0.141 &  6.896 \\ \hline
& SR3 & 21.596 & 25.587 & 0.683 & 0.231 & 0.299 & 2.065 & 0.357 & 6.649 \\
Diffusion-based & LDM &  24.234 & 29.655 & 0.780 &  0.122 &  0.244 &  0.898 &  0.185 &  5.794 \\ 
& IDM &  24.573 & 29.526 & 0.716 & 0.149 & 0.294 & 0.651 & 0.227& 6.496 \\
& LDM-SS &  26.047 & 31.447 & 0.823 &  0.141 &  0.227 &  1.120 &  0.194 &  5.195 \\
\bottomrule
\end{tabular}
\end{adjustbox}
\caption{Performance comparison of $\times$4 (32x32 $\rightarrow$ 128x128) SR models on DIV2K validation set. Even though LDM-SS successfully suppresses distortions, contributing to visually enhanced and reliable outputs, there exists an intriguing divergence between the notable visual performance and the quantitative measures.}
\label{table:fusion_results}
\end{center}  
\vspace{-10pt}
\end{table*}

\subsection{Comparison with the State-of-the-Art Methods}
\textbf{Quantitative Comparison.}
Table \ref{table:fusion_results} demonstrates quantitative comparison for 4$\times$ SR methods and our proposed strategic sample selection approach for Task-2. The main objective of Task-2 is to have the subjects concentrate exclusively on choosing the image that appears more "photo-realistic" among 15 diffusion samples. In each round, fifteen 128x128 samples are presented simultaneously to each participant and asked to select at most 3 images with natural looking details, colors and lightning. To have a concrete evaluation, this selection process is repeated for 10 images from DIV2K  \cite{Agustsson_2017_CVPR_Workshops} dataset. Also, in order not to biased the participants, the ground-truth image is kept hidden. The top 3 selected images are ensembled by pixel-wise averaging. Then, we compare the ensembled images with the existing state-of-the-art methods including EDSR \cite{EDSR2017}, RRDB \cite{wang2018esrgan}, HAT \cite{hat_chen2023activating}, ESRGAN+ \cite{esrganplus}, SPSR \cite{ma_SPSR}, LDL \cite{details_or_artifacts}, SROOE \cite{srooe_Park_2023_CVPR} as well as stochastic SR methods like HCFLow++ \cite{hcflow_liang21hierarchical}, SRFlow-DA \cite{jo2021srflowda}. Even though we provide quantitative comparison results for our proposed approach, the efficacy of evaluating visual artifacts in SR tasks cannot solely rely on metrics such as PSNR or other quantitative perceptual scores. While these metrics provide numerical insights into image quality, they might not capture the nuances of visual artifacts effectively. In this context, it becomes crucial to explore alternative methods that go beyond quantitative assessments.
\vspace{12pt}

\noindent
\textbf{Qualitative Comparison.}
Visual comparisons among 4$\times$ SR approaches and LDM-SS are presented in Fig. \ref{fig:first_img}, \ref{fig:mnist} and \ref{fig:visuals_div2k}. These figures showcase the effectiveness of strategic sample selection by human feedback in mitigating visual artifacts. In details, we observe that
all GAN-SR results including ESRGAN+ \cite{esrganplus}, SPSR \cite{ma_SPSR}, LDL \cite{details_or_artifacts}, SROOE \cite{srooe_Park_2023_CVPR} as well as stochastic SR methods like HCFLow++ \cite{hcflow_liang21hierarchical}, SRFlow-DA \cite{jo2021srflowda} produce visible artifacts and experience excessive sharpness. For instance, the visual results presented in Fig. \ref{fig:mnist} demonstrate the effectiveness of human feedback for information-centric applications by enabling the outcome of the accurate digit. The first digit presented by state-of-the-art methods can be perceived as ``6." On the contrary, LDM-SS provides the accurate digit ``5." Similarly, second digit obtained by SOTA methods is not a clear ``8", unlike the output of LDM-SS. In addition, the qualitative results from DIV2K \cite{Agustsson_2017_CVPR_Workshops} dataset demonstrated in \ref{fig:visuals_div2k} prove LDM-SS visibly suppresses unwanted distortions and enhances overall image quality in a perceptually meaningful way and enables visually on-par or better results with SOTA SR methods.
\vspace{-4pt}

\section{Conclusion}
\vspace{-4pt}
DMs are able to generate not one but a set of plausible SR images at their output. While this improves diversity, it brings the ambiguity of how to select the single trustworthy SR solution when the goal is to extract  crucial information from LR~images. In this work, we are primarily interested in obtaining a consistent and reliable image SR result within the space spanned by diffusion models. Specifically, we  benefit from human feedback while selecting diverse set of diffusion samples since we found that we cannot rely on quantitative metrics to select a trustworthy result. Then, we ensemble the selected images to resolve the ambiguity in SR images and to obtain a reliable, photo-realistic solution. Our approach achieves promising results on DIV2K validation set and information-centric applications. While our method, LDM-SS, excels in delivering visually appealing results by reducing artifacts, these improvements may not be accurately reflected in quantitative scores like PSNR, MS-SSIM, LPIPS, DISTS, etc.  The proposed method for selection of diffusion samples is generic in the sense that any off-the-shelf diffusion model can be easily plugged into this framework to benefit from human feedback to resolve the ambiguity.


\small
\bibliographystyle{IEEEbib}
\bibliography{source}

\begin{thebibliography}{10}

\bibitem{EDSR2017}
B.~Lim, S.~Son, H.~Kim, S.~Nah, and K.~Lee,
\newblock ``Enhanced deep residual networks for single image super-resolution,''
\newblock in {\em IEEE/CVF CVPR Workshops}, 2017.

\bibitem{esrganplus}
N.~Rakotonirina and A.~Rasoanaivo,
\newblock ``Esrgan+: Further improving enhanced super-resolution generative adversarial network,''
\newblock in {\em IEEE ICASSP}, 2020, pp. 3637--3641.

\bibitem{srooe_Park_2023_CVPR}
S.~Park, Y.~Moon, and N.~Cho,
\newblock ``Perception-oriented single image super-resolution using optimal objective estimation,''
\newblock in {\em IEEE/CVF Conf. on Comp. Vision and Patt. Recog. (CVPR)}, 2023, pp. 1725--1735.

\bibitem{wang2018esrgan}
X.~Wang, Ke~Yu, S.~Wu, J.~Gu, Y.~Liu, C.~Dong, Y.~Qiao, and C.~C.~Loy,
\newblock ``{ESRGAN:} enhanced super-resolution generative adversarial networks,''
\newblock in {\em European Conf. on Comp. Vision (ECCV) Workshops}, 2018.

\bibitem{ma_SPSR}
C.~Ma, Y.~Rao, Y.~Cheng, C.~Chen, J.~Lu, and J.~Zhou,
\newblock ``Structure-preserving super resolution with gradient guidance,''
\newblock in {\em IEEE/CVF Conf. Comp. Vis. and Patt. Recog. (CVPR)}, 2020.

\bibitem{hat_chen2023activating}
X.~Chen, X.~Wang, J.~Zhou, Y.~Qiao, and C.~Dong,
\newblock ``Activating more pixels in image super-resolution transformer,''
\newblock in {\em IEEE/CVF Conf. Comp. Vision and Patt. Recog. (CVPR)}, 2023.

\bibitem{details_or_artifacts}
J.~Liang, H.~Zeng, and L.~Zhang,
\newblock ``Details or artifacts: A locally discriminative learning approach to realistic image super-resolution,''
\newblock in {\em IEEE/CVF Conf. on Comp. Vision and Patt. Recog. (CVPR)}, 2022, pp. 5657--5666.

\bibitem{urban100_cite}
J.~Huang, A.~Singh, and N.~Ahuja,
\newblock ``Single image super-resolution from transformed self-exemplars,''
\newblock in {\em IEEE Conf. on Comp. Vision and Patt. Recog. (CVPR)}, 2015.

\bibitem{dong2015image}
C.~Dong, Loy C., K.~He, and X.~Tang,
\newblock ``Image super-resolution using deep convolutional networks,''
\newblock {\em IEEE Trans. on Pattern Analysis and Mach. Intell.}, vol. 38, pp. 295--307, 2016.

\bibitem{kim2016accurate}
J.~Kim, J.~Kwon~Lee, and K.~Mu~Lee,
\newblock ``Accurate image super-resolution using very deep convolutional networks,''
\newblock {\em IEEE/CVF Conf. on Comp. Vision and Patt. Recog. (CVPR)}, pp. 1646--1654, 2016.

\bibitem{ledig2017photorealistic}
C.~Ledig, L.~Theis, F.~Husz{\'a}r, J.~Caballero, A.~Aitken, A.~Tejani, J.~Totz, Z.~Wang, and W.~Shi,
\newblock ``Photo-realistic single image super-resolution using a generative adversarial network,''
\newblock {\em IEEE/CVF Conf. on Comp. Vision and Patt. Recog. (CVPR)}, pp. 105--114, 2017.

\bibitem{srcnn}
G.~Lin, Q.~Wu, X.~Huang, Q.~Lida, and X.~Chen,
\newblock ``Deep convolutional networks-based image super-resolution,''
\newblock in {\em Int. Conf. Intelligent Computing}, 2017, pp. 338--344.

\bibitem{RCAN2018}
Y.~Zhang, K.~Li, K.~Li, L.~Wang, B.~Zhong, and Y.~Fu,
\newblock ``Image super-resolution using very deep residual channel attention networks,''
\newblock in {\em IEEE/CVF ECCV}, 2018.

\bibitem{lugmayr2020srflow}
L.~Andreas, D.~Martin, L.~Van~Gool, and R.~Timofte,
\newblock ``Srflow: Learning the super-resolution space with normalizing flow,''
\newblock in {\em ECCV}, 2020.

\bibitem{jo2021srflowda}
Y.~Jo, S.~Yang, and S.~Joo Kim,
\newblock ``Srflow-da: Super-resolution using normalizing flow with deep convolutional block,''
\newblock in {\em IEEE/CVF Conf. on Comp. Vision and Patt. Recog. (CVPR) Workshops}, June 2021.

\bibitem{SR3_saharia2022image}
C.~Saharia, J.~Ho, W.~Chan, T.~Salimans, D.~Fleet, and M.~Norouzi,
\newblock ``Image super-resolution via iterative refinement,''
\newblock {\em IEEE Transactions on Pattern Analysis and Machine Intelligence}, vol. 45, no. 4, pp. 4713--4726, 2022.

\bibitem{LDM_rombach2022high}
R.~Rombach, A.~Blattmann, D.~Lorenz, P.~Esser, and B.~Ommer,
\newblock ``High-resolution image synthesis with latent diffusion models,''
\newblock in {\em IEEE/CVF Conf. on Comp. Vision and Patt. Recog. (CVPR)}, 2022, pp. 10684--10695.

\bibitem{korkmaz2022perception}
C.~Korkmaz, A.~M. Tekalp, Z.~Do{\u{g}}an, E.~Erdem, and A.~Erdem,
\newblock ``Perception-distortion trade-off in the sr space spanned by flow models,''
\newblock in {\em IEEE Int. Cong. on Image Processing (ICIP)}. IEEE, 2022, pp. 2396--2400.

\bibitem{diff_mixture_of_experts_luo2023image}
F.~Luo, J.~Xiang, J.~Zhang, X.~Han, and W.~Yang,
\newblock ``Image super-resolution via latent diffusion: A sampling-space mixture of experts and frequency-augmented decoder approach,''
\newblock {\em arXiv preprint arXiv:2310.12004}, 2023.

\bibitem{learning_SR_space_2021}
Andreas Lugmayr and et. al.,
\newblock ``Ntire 2021 learning the super-resolution space challenge,''
\newblock in {\em IEEE Conf. on Comp. Vision and Patt. Recog. Workshops (CVPRW)}, 2021, pp. 596--612.

\bibitem{learning_SR_space_2022}
Andreas Lugmayr and et. al.,
\newblock ``Ntire 2022 challenge on learning the super-resolution space,''
\newblock in {\em IEEE Conf. on Comp. Vision and Patt. Recog. Workshops (CVPRW)}, 2022, pp. 785--796.

\bibitem{lpips}
R.~Zhang, P.~Isola, A.~Efros, E.~Shechtman, and O.~Wang,
\newblock ``The unreasonable effectiveness of deep features as a perceptual metric,''
\newblock in {\em IEEE/CVF Conf. Comp. Vision and Patt. Recog. (CVPR)}, 2018, pp. 586--595.

\bibitem{dists}
K.~Ding, K.~Ma, S.~Wang, and E.~Simoncelli,
\newblock ``Image quality assessment: Unifying structure and texture similarity,''
\newblock {\em IEEE Trans. Patt Anal. Mach. Intel.}, vol. 44, pp. 2567--2581, 2020.

\bibitem{PDASR}
Y.~Zhang, B.~Ji, J.~Hao, and A.~Yao,
\newblock ``Perception-distortion balanced admm optimization for single-image super-resolution,''
\newblock in {\em European Conf. on Comp. Vision (ECCV)}, 2022.

\bibitem{ian_gan}
I.~Goodfellow, J.~Pouget-Abadie, M.~Mirza, B.~Xu, D.~Warde-Farley, S.~Ozair, A.~Courville, and Y.~Bengio,
\newblock ``Generative adversarial nets,''
\newblock in {\em Advances in Neural Information Processing Systems}, 2014, vol.~27.

\bibitem{vae_liu2021variational}
Z.~Liu, W.~Siu, and L.~Wang,
\newblock ``Variational autoencoder for reference based image super-resolution,''
\newblock in {\em IEEE/CVF Conf. on Comp. Vision and Patt. Recog. (CVPR)}, 2021, pp. 516--525.

\bibitem{vae_zhou2021vspsr}
H.~Zhou, C.~Huang, S.~Gao, and X.~Zhuang,
\newblock ``Vspsr: Explorable super-resolution via variational sparse representation,''
\newblock in {\em IEEE/CVF Conf. on Computer Vision and Pattern Recognition}, 2021, pp. 373--381.

\bibitem{fsncsr_song2022fs}
K.~Song, D.~Shim, K.~Kim, J.~Lee, and Y.~Kim,
\newblock ``{FS-NCSR}: Increasing diversity of the super-resolution space via frequency separation and noise-conditioned normalizing flow,''
\newblock in {\em IEEE/CVF Conf. Comp. Vis. and Patt. Recog. Workshops (CVPRW)}, 2022, pp. 967--976.

\bibitem{arm_van2016conditional}
A.~Van~den Oord, N.~Kalchbrenner, L.~Espeholt, O.~Vinyals, A.~Graves, et~al.,
\newblock ``Conditional image generation with pixelcnn decoders,''
\newblock {\em Advances in neural information processing systems}, vol. 29, 2016.

\bibitem{arm_pmlr-v119-chen20s}
M.~Chen, A.~Radford, R.~Child, J.~Wu, H.~Jun, D.~Luan, and I.~Sutskever,
\newblock ``Generative pretraining from pixels,''
\newblock in {\em Int. Conf. on Mach. Learning}, 2020, pp. 1691--1703.

\bibitem{arm_salimans2017pixelcnn++}
T.~Salimans, A.~Karpathy, X.~Chen, and D.~Kingma,
\newblock ``Pixelcnn++: Improving the pixelcnn with discretized logistic mixture likelihood and other modifications,''
\newblock in {\em Int. Conf. on Learning Representations}, 2016.

\bibitem{diff_NEURIPS2020_4c5bcfec}
J.~Ho, A.~Jain, and P.~Abbeel,
\newblock ``Denoising diffusion probabilistic models,''
\newblock in {\em Advances in Neural Information Processing Systems}, 2020.

\bibitem{diff_implicit_chen2021learning}
Y.~Chen, S.~Liu, and X.~Wang,
\newblock ``Learning continuous image representation with local implicit image function,''
\newblock in {\em IEEE/CVF conference on computer vision and pattern recognition}, 2021, pp. 8628--8638.

\bibitem{diff_avrahami2022blended}
O.~Avrahami, D.~Lischinski, and O.~Fried,
\newblock ``Blended diffusion for text-driven editing of natural images,''
\newblock in {\em IEEE/CVF Conf. Comp. Vis. and Patt. Recog. (CVPR)}, 2022, pp. 18208--18218.

\bibitem{diff_gu2022vector}
S.~Gu, D.~Chen, J.~Bao, F.~Wen, B.~Zhang, D.~Chen, L.~Yuan, and B.~Guo,
\newblock ``Vector quantized diffusion model for text-to-image synthesis,''
\newblock in {\em IEEE/CVF Conf. on Comp. Vision and Pattern Recog. (CVPR)}, 2022, pp. 10696--10706.

\bibitem{diffretrival_NEURIPS2022_62868cc2}
A.~Blattmann, R.~Rombach, K.~Oktay, J.~M\"{u}ller, and B.~Ommer,
\newblock ``Retrieval-augmented diffusion models,''
\newblock in {\em Advances in Neural Info. Processing Systems}, 2022, pp. 15309--15324.

\bibitem{hcflow_liang21hierarchical}
J.~Liang, A.~Lugmayr, K.~Zhang, M.~Danelljan, L.~Van~Gool, and R.~Timofte,
\newblock ``Hierarchical conditional flow: A unified framework for image super-resolution and image rescaling,''
\newblock in {\em IEEE Int. Conf. on Computer Vision}, 2021.

\bibitem{deng2012mnist}
Li~Deng,
\newblock ``The mnist database of handwritten digit images for machine learning research,''
\newblock {\em IEEE Signal Processing Magazine}, vol. 29, no. 6, pp. 141--142, 2012.

\bibitem{Agustsson_2017_CVPR_Workshops}
E~Agustsson and R.~Timofte,
\newblock ``{NTIRE 2017 Challenge} on single image super-resolution: Dataset and study,''
\newblock in {\em IEEE/CVF Conf. Comp. Vis. and Patt. Recog. (CVPR) Workshops}, 2017.

\bibitem{implicit_gao2023implicit}
S.~Gao, X.~Liu, B.~Zeng, S.~Xu, Y.~Li, X.~Luo, J.~Liu, X.~Zhen, and B.~Zhang,
\newblock ``Implicit diffusion models for continuous super-resolution,''
\newblock in {\em IEEE/CVF Conf. Comp. Vis. and Patt. Recog.}, 2023, pp. 10021--10030.

\end{thebibliography}

\end{document}